\documentclass[prc,twocolumn,showpacs,preprintnumbers,amsmath,amssymb,superscriptaddress]{revtex4}
\usepackage{amsfonts}

\usepackage{graphicx} % Include figure files
\usepackage{dcolumn}  % Align table columns on decimal point
\usepackage{bm}       % bold math
\newcommand{\hfbax}{\sc hfb-ax}

\newcommand{\hfbrad}{\sc hfbrad}

\newcommand{\rr} {\boldsymbol{r}}
\newcommand{\rp} {\boldsymbol{p}}
\newcommand{\bea}{\begin{eqnarray}}
\newcommand{\eea}{\end{eqnarray}}

\begin{document}
%
%\ifx \fi

\title{Quasi-particle continuum and resonances in the Hartree-Fock-Bogoliubov theory}

\author{J.C. Pei}
%\email{peij@ornl.gov}
 \affiliation{Joint Institute for Heavy Ion
Research, Oak Ridge, TN 37831}
 \affiliation{Department of Physics and
Astronomy, University of Tennessee Knoxville, TN 37996}
\affiliation{Physics Division, Oak Ridge National Laboratory, P.O.
Box 2008, Oak Ridge, TN 37831}

\author{A.T. Kruppa}
%\email{atk@namafia.atomki.hu}
 \affiliation{Joint Institute for Heavy Ion
Research, Oak Ridge, TN 37831}
 \affiliation{Department of Physics and
Astronomy, University of Tennessee Knoxville, TN 37996}
\affiliation{Institute of Nuclear Research,
  P.O. Box 51, H-4001 Debrecen, Hungary}

\author{W. Nazarewicz}
%\email{witek@utk.edu}
 \affiliation{Department of Physics and
Astronomy, University of Tennessee Knoxville, TN 37996}
\affiliation{Physics Division, Oak Ridge National Laboratory, P.O.
Box 2008, Oak Ridge, TN 37831} \affiliation{Institute of Theoretical
Physics, Warsaw University, ul. Ho\.{z}a 69, PL-00681 Warsaw,
Poland}

\date{\today}

\begin{abstract}
The quasi-particle energy spectrum of the
Hartree-Fock-Bogoliubov (HFB) equations contains discrete bound
states, resonances, and  non-resonant continuum states. We study the structure of
the unbound quasi-particle spectrum of  weakly bound nuclei within several
methods that do not rely on imposing scattering or outgoing boundary conditions.
Various approximations  are examined to estimate resonance widths.
It is shown that the stabilization method
works well for all HFB resonances except for very narrow ones.
 The Thomas-Fermi approximation to the non-resonant continuum has been shown
 to be very effective, especially for coordinate-space HFB calculations in
 large boxes that involve huge amounts of discretized quasi-particle continuum states.

\end{abstract}


\maketitle

\section{Introduction}

A major challenge for theoretical nuclear structure research is the
development of robust models and techniques aiming at the microscopic
description of the nuclear many-body  problem and capable of extrapolating
into unknown regions of the nuclear landscape. Since the coherent theoretical
framework should describe both well-bound and drip-line nuclei,
of particular importance is the treatment of the particle continuum in
weakly bound nuclei and the development of microscopic reaction theory that is integrated with improved structure models \cite{[Dob07b]}.

For open-shell medium-mass and heavy nuclei, the theoretical tool of choice is the nuclear
Density Functional Theory (DFT) \cite{[Ben03]}. Its main ingredient is the energy density functional that depends on proton
and neutron densities and currents, as well as pairing densities describing nuclear superconductivity \cite{[Roh10]}. The  HFB  equations of nuclear DFT
properly take into account the scattering continuum \cite{bulgac80,[Dob84],Belyaev,dobaczewski96}, and this is a welcome feature of the formalism.

The quasi-particle energy spectrum of HFB consists of a finite number of
bound quasi-particle states and a continuum of unbound states.
The so-called deep-hole quasi-particle resonances are unique to the HFB theory \cite{Belyaev}. They can be associated with
single-particle Hartree-Fock (HF) states that are well bound in the absence of pairing correlations but acquire a finite particle width due to the continuum
coupling induced by the particle-particle channel. As discussed in the literature \cite{dobaczewski96,Michel}, those resonances cannot be properly described by BCS-like theories
\cite{betan,[Bet08]} which yield particle and pairing densities which are not  localized in space.

The role played by the HFB continuum has been
a subject of  many works \cite{bulgac80,[Dob84],Belyaev,[Fay98],dobaczewski96,[Dob01a],Grasso,Grasso1,[Bul02],[Yu03],[Bor06],teran,schunck08,oba,zhang}. The proper
treatment of quasi-particle continuum  is important
not only for ground-state  properties but also for the description of
nuclear excitations, e.g.,  within a self-consistent QRPA
approach \cite{rodin,Hagino,[Ter06],[Ter06a],Mizuyama}.  Within the real-energy HFB framework, the proper
theoretical treatment of the HFB continuum is fairly  sophisticated
since the scattering boundary conditions must be met.

If the outgoing boundary conditions are imposed, the  unbound HFB eigenstates
have complex energies; their imaginary parts are related to the particle width. The complex-energy spherical HFB equations have
been solved in Ref.~\cite{Michel} within the Gamow HFB (GHFB)
approach, which shares many techniques with
the Gamow shell model \cite{csm,michel09}. Within GHFB,  quasi-particle resonance widths can be calculated
with a high precision. Alternatively, diagonalizing the HFB matrix in the Gamow HF or   P\"oschl-Teller-Ginocchio basis turned out to be an efficient way to
account for the continuum effects \cite{Stoitsov08}.

In addition to the methods
that  employ correct asymptotic boundary conditions for unbound states, the quasi-particle continuum
can be approximately treated by means of a discretization method.  The commonly used approach is to impose  the box boundary
conditions \cite{dobaczewski96,HFBRAD,[Yam05a],[Pei08],Oberacker,teran,pei09}, in which
wave functions are spanned by a basis of orthonormal functions
defined on a lattice in  coordinate space and
enforced to be zero at box boundaries. In this way, referred to as the  $\mathcal{L}^2$ discretization, quasi-particle continuum of HFB  is
represented by a  finite number of box  states. It has been demonstrated by explicit calculations for   weakly bound nuclei \cite{Michel,Grasso} that such a box discretization is very accurate  when compared to the exact results.

In many practical applications involving complex geometries of nucleonic densities in two or three dimensions, such as those appearing in nuclear fission and fusion  or weakly bound and spatially-extended systems, it is crucial to consider large coordinate spaces. At the same time, employed  lattice spacings should be sufficiently small to provide good resolution and numerical accuracy. As a result, the size of the discretized continuum space may often become intractable. This is also the case for calculations employing multiresolution pseudo-spectral methods \cite{gif} which effectively invoke enormous  continuum spaces.
Therefore, it becomes exceedingly important  to develop methods allowing precise treatment of
HFB resonances and non-resonant quasi-particle continuum without resorting  to  the explicit computation of all box states. Our paper is devoted to  this problem. Namely, we study  the effect of high-energy, non-resonant quasi-particle continuum
on HFB equations and observables and propose an efficient  scheme
to account for those states. The technique, based on the
local-density approximation to  HFB-Popov equations for Bose
gases \cite{Reidl}, was previously used to approximate the
continuum contribution in solving the Bogoliubov de-Gennes equations
for cold Fermi gases \cite{liu}. Here, we extend this method to the nuclear Skyrme HFB approach.

We also study  quasi-particle  resonances and devise techniques to isolate them and estimate their  widths using   $\mathcal{L}^2$ discretization.
Several approximate methods are  examined. In atomic physics, the so-called
stabilization method has been widely used to precisely calculate
resonance widths \cite{hazi,ryaboy,mandelshtam94,kruppa99}. A
modified stabilization method, based on box
solutions, was  developed \cite{mandelshtam94} and  used   to
study single-particle resonances in nuclei \cite{Giai04}. Here, we  demonstrate that the stabilization method also works
reliably for quasi-particle resonances of HFB. Besides the stabilization
method, a straightforward smoothing and fitting method that utilizes the density of box states is proposed and tested. Finally, we assess the quality of the perturbative expression \cite{Belyaev} for
deep-hole resonance widths.

This paper is organized as follows. Section~\ref{HFBbox} briefly reviews the properties of HFB equations. In Sec.~\ref{HFBlocal}, we apply the
local-density approximation to account for the high-energy quasi-particle continuum. We  propose and test a hybrid HFB strategy that makes it possible to solve HFB equations assuming a  low energy cutoff, thus appreciably reducing the computational effort. Section~\ref{HFBres}
studies deep-hole HFB resonances
with several  methods: the smoothing and fitting method,
the box stabilization method, and the perturbation expression. The
applicability of each technique is examined. Section~\ref{examples}
 contains illustrative examples for weakly bound Ni
isotopes. Finally, conclusions are given in Sec.~\ref{conclusions}.

\section{Coordinate-space HFB formalism}\label{HFBbox}

In this section, we briefly
recall general properties of the HFB eigenstates. The  HFB equation in the
coordinate space can be written as:
\bea
  \left[
\begin{array}{cccc}%
 \displaystyle h-\lambda& {\hspace{0.7cm} } \tilde{h} \vspace{2pt}\\
  \displaystyle \tilde{h}& -h+\lambda \\
\end{array}
\right]\left[
\begin{array}{clrr}%
u_i(\rr) \\ \vspace{2pt} v_i(\rr)\\
\end{array}
\right]=E_i\left[
\begin{array}{clrr}%
 u_i(\rr)  \vspace{2pt}\\ v_i(\rr)
\end{array}
\right],\label{HFB}
\eea
where $h$ is the HF Hamiltonian; $\tilde{h}$ is the pairing Hamiltonian
as defined in Ref.~\cite{dobaczewski96};
$u_i$ and $v_i$ are the upper and lower components of quasi-particle
wave functions, respectively; $E_i$ is the quasi-particle energy; and
$\lambda$ is the Fermi energy (or chemical potential).
For bound systems, $\lambda < 0$ and the self-consistent densities and fields
are localized in space. In our work, we shall limit discussion to local mean fields. To relate better to other papers, in the following we should use the notation
$\Delta(\rr)= \tilde{h}(\rr)$ for the pairing potential.

For $|E_i|<-\lambda$, the eigenstates of (\ref{HFB}) are discrete and
$v_i(\rr)$ and $u_i(\rr)$ decay exponentially.
The quasi-particle continuum corresponds to  $|E_k|>-\lambda$. For those states, the upper component of the
wave function always has  a scattering asymptotic form. By applying the
box boundary condition, the continuum becomes discretized and one obtains
a finite number of continuum quasi-particles. In principle, the box solution
representing the continuum can be  close to the exact
solution when a sufficiently big box and small mesh size are adopted.

The solution of Eq.~(\ref{HFB}) in non-spherical boxes is a difficult
computational task. The recently developed parallel
2D-HFB solvers utilizing the B-spline technique offer excellent
accuracy when describing weakly bound nuclei and large
deformations~\cite{teran,[Pei08]}. To solve the HFB equations in a
3D coordinate space is more complicated, but the development of
a general-purpose 3D-HFB solver based on multiresolution analysis
and wavelet expansion~\cite{gif,[Pei08]} is underway. With the
2D-HFB box solver {\hfbax} \cite{[Pei08]}, the HFB equation is
solved by discretizing wave functions on a 2D lattice.
The discretization precision depends on the order of B-splines, the
maximum mesh size, and the box size. Using this solver, one can obtain an
extremely dense quasi-particle energy spectrum by adopting a large
box size. For example, there are about 7,000 states below 60 MeV if
a square box of 40$\times$40 fm is used. In this case,  the 2D box solution corresponds to a high resolution of discretized continuum states.

\section{Thomas-Fermi approximation to high energy HFB continuum}\label{HFBlocal}

In this section, the Thomas-Fermi (TF)
approximation \cite{Reidl,liu} is applied to study   high-energy HFB continuum contributions. Calculations are carried out within  the Skyrme HFB approach.
The accuracy of this method is
tested by comparing with the results obtained using the full box discretization method.

\subsection{Contribution from high-energy continuum states}

The TF approximation to the HFB continuum  was originally
applied in the context of 1D Fermi gases \cite{liu}. Following the method of
\cite{Reidl,liu}, we derive expressions for the continuum components to
nucleonic densities due to
high-energy scattering states within the Skyrme HFB approach.

The high-energy HFB wave functions $u(\rr)$ and
$v(\rr)$ can be  approximated by
\cite{Reidl,[Bul02],[Yu03],[Bor06]}:
\begin{equation}
u(\rr)\rightarrow u(\rp,\rr)e^{i\hbar\phi(\rr)},
~~v(\rr)\rightarrow v(\rp,\rr)e^{i\hbar\phi(\rr)},
\end{equation}
where ${\boldsymbol{\nabla}}\phi(\rr)\equiv\rp$.
In this
approximation, the derivatives of $u(\rp,\rr)$ and $v(\rp,\rr)$, and the second
derivatives of $\phi(\rr)$ are neglected, as well as the derivative terms of
the effective mass and spin-orbit.
The latter terms have negligible effects on high-energy states, and they are usually excluded in the pairing regularization
procedure~\cite{[Bul02],[Yu03],[Bor06]}.
The corresponding HFB equation (\ref{HFB}) can be written  as
\begin{subequations}
\begin{equation}
\begin{array}{l}
~~~\displaystyle\left(\frac{\hbar^2p^2}{2M^{*}}+V(\rr)-\lambda\right)u(\rp,
\rr)+\Delta(\rr) v(\rp, \rr) \vspace{5pt}\\~~~~~~~ = E(\rp,
\rr) u(\rp, \rr),
\end{array}
\end{equation}
\begin{equation}
\begin{array}{l}
~~~\displaystyle\left(\frac{\hbar^2p^2}{2M^{*}}+V(\rr)-\lambda\right)v(\rp,
\rr)-\Delta(\rr) u(\rp, \rr) \vspace{5pt}\\~~~~~~~ = -E(\rp,
\rr) v(\rp, \rr),
\end{array}
\end{equation}
\end{subequations}
where $V(\rr)$ is the Skyrme HF potential; $M^{*}(\rr)$ is
the density-dependent effective mass; and $E(\rp, \rr)$ denotes
the local quasi-particle energy. By introducing
the  HF energy $\varepsilon_{\rm HF}(\rp, \rr)$ relative to the chemical potential,
\begin{equation}\label{eHF}
\varepsilon_{\rm HF}(\rp,\rr) =
\frac{\hbar^2\rp^2}{2M^{*}(\rr)}+V(\rr)-\lambda,
\end{equation}
$E(\rp, \rr)$ takes the familiar BCS-like form:
\begin{equation}
E(\rp, \rr) = \sqrt{\varepsilon_{\rm
HF}^2(\rp,\rr)+\Delta^2(\rr)}.
\end{equation}

In this work we consider the contact pairing interaction with the density-dependent pairing strength $V_{pair}(\rr)$. The resulting
 pairing potential $\Delta(\rr)$ can be written as:
\begin{equation}{\label{pairpot}}
\begin{array}{l}
\Delta(\rr) = \displaystyle V_{pair}(\rr)\tilde{\rho}(\rr)= {1\over 2}  V_0\left[1-\frac{\rho(\rr)}{2\rho_0}\right]\tilde{\rho}(\rr), \vspace{6pt}\\
\tilde{\rho}(\rr) =  \displaystyle -\sum_i u_i(\rr)v_i^{*}(\rr),
\end{array}
\end{equation}
where $V_{0}$ is the pairing interaction strength; $\rho_0$ is the  saturation density 0.16 fm$^{-3}$; and
$\rho(\rr)$ and $\tilde{\rho}(\rr)$ are the particle density and pairing density, respectively. The pairing potential (\ref{pairpot}) corresponds to the so-called   mixed pairing interaction \cite{[Dob02c]}.

The normalization condition in the $\rp$ space,  $|u(\rp,\rr)|^2+|v(\rp,\rr)|^2=1$,
implies that \cite{liu}
\begin{subequations}
\begin{equation}
v^2_p = \frac{1}{2}\left(1-\frac{\varepsilon_{\rm HF}(\rp,\rr)}{E(\rp,\rr)}\right),
\end{equation}
\begin{equation}
u^2_p = \frac{1}{2}\left(1+\frac{\varepsilon_{\rm HF}(\rp,\rr)}{E(\rp,\rr)}\right).
\end{equation}
\end{subequations}
Consequently, the contribution to the particle density from the high-energy continuum states is  given by
\begin{equation}{\label{densit-c}}
\rho_c(\rr) = \int\hspace{-1.4em}\sum_p\left(1-\frac{\varepsilon_{\rm
HF}}{E}\right)\Theta[E-E_c],
\end{equation}
where $E_c$ is the  quasi-particle energy cutoff above which the TF
approximation is applied.  In Eq.~(\ref{densit-c}), the sum/integral symbol
denotes the summation over the discretized continuum box states or, alternatively, the  integration in the momentum space if the HFB
equations are solved with the outgoing or scattering boundary conditions.
Similarly, the high-energy continuum contributions
to the kinetic density, pairing density, and pairing potential are given respectively by:
\begin{eqnarray}
\tau_c(\rr) & = & \int\hspace{-1.4em}\sum_p p^2\left(1-\frac{\varepsilon_{\rm
HF}}{E}\right)\Theta[E-E_c], \label{kinetic-c}\\
\tilde{\rho}_c(\rr)& = & -\int\hspace{-1.4em}\sum_p v_pu_p = -\int\hspace{-1.4em}\sum_p
\frac{\Delta}{E}\Theta[E-E_c],\label{paird-c} \\
\Delta_c(\rr) & = & V_{pair}(\rr)\tilde{\rho}_c(\rr) \nonumber\\
& =& -V_{pair}(\rr)\int\hspace{-1.4em}\sum_p \frac{\Delta}{E}\Theta[E-E_c] \label{pairD}.
\end{eqnarray}
By separating the continuum contribution from the equation (\ref{pairpot}), we see that the TF procedure \cite{Reidl,liu}  is formally equivalent to the
pairing regularization scheme with an effective pairing interaction $V_{eff}(\rr)$ \cite{[Bul02],[Yu03],[Bor06]}:
\begin{equation}{\label{effp-c}}
\frac{1}{V_{eff}(\rr)}= \frac{1}{V_{pair}(\rr)} - \sum_p
\frac{1}{E}\Theta[E-E_c].
\end{equation}

The expressions (\ref{densit-c}-\ref{pairD}) can be written in a compact form by replacing the momentum sum/integral  by an integral over quasi-particle
energy space between $E_c$ and the maximum cutoff energy considered $E_{m}$:
\begin{eqnarray}\label{cdensit}
\rho_c(\rr)  & = & \frac{{M^*(\rr)}}{2\pi^2\hbar^2}
\int_{E_c}^{E_{m}}\!\!\! dE \left(\frac{E}
{\sqrt{E^2-\Delta^2(\rr)}}-1\right) p(E,\rr), \nonumber \\
\tau_c(\rr) &=&  \frac{{M^*(\rr)}}{2\pi^2\hbar^2}
\int_{E_c}^{E_{m}} \!\!\!dE
\left(\frac{E}{\sqrt{E^2-\Delta^2(\rr)}}-1\right) p^3(E,\rr), \nonumber \\
\tilde{\rho}_c(\rr) & =&
-\frac{{M^*(\rr)\Delta(\rr)}}{2\pi^2\hbar^2}\int_{E_c}^{E_{m}}\!
\frac{p(E,\rr)\,dE}{\sqrt{E^2-\Delta^2(\rr)}},
\end{eqnarray}
where
\begin{equation}
p(E,\rr)\equiv \sqrt
{\frac{2M^{*}(\rr)}{\hbar^2}\left[\sqrt{E^2-\Delta^2(\rr)} - V(\rr) +
\lambda \right]}.
\end{equation}
The effective pairing strength (\ref{effp-c}) becomes
\begin{equation}{\label{pair-reg}}
 \frac{1}{V_{eff}(\rr)} =
\frac{1}{V_{pair}(\rr)}-\frac{{M^{*}(\rr)}}{2\pi^2\hbar^2}\int_{E_c}^{E_{m}}\!
\frac{p(E,\rr)\,dE}{\sqrt{E^2-\Delta^2(\rr)}}.
\end{equation}

\begin{figure}[htb]
\centerline{\includegraphics[trim=0cm 0cm 0cm
0cm,width=0.44\textwidth,clip]{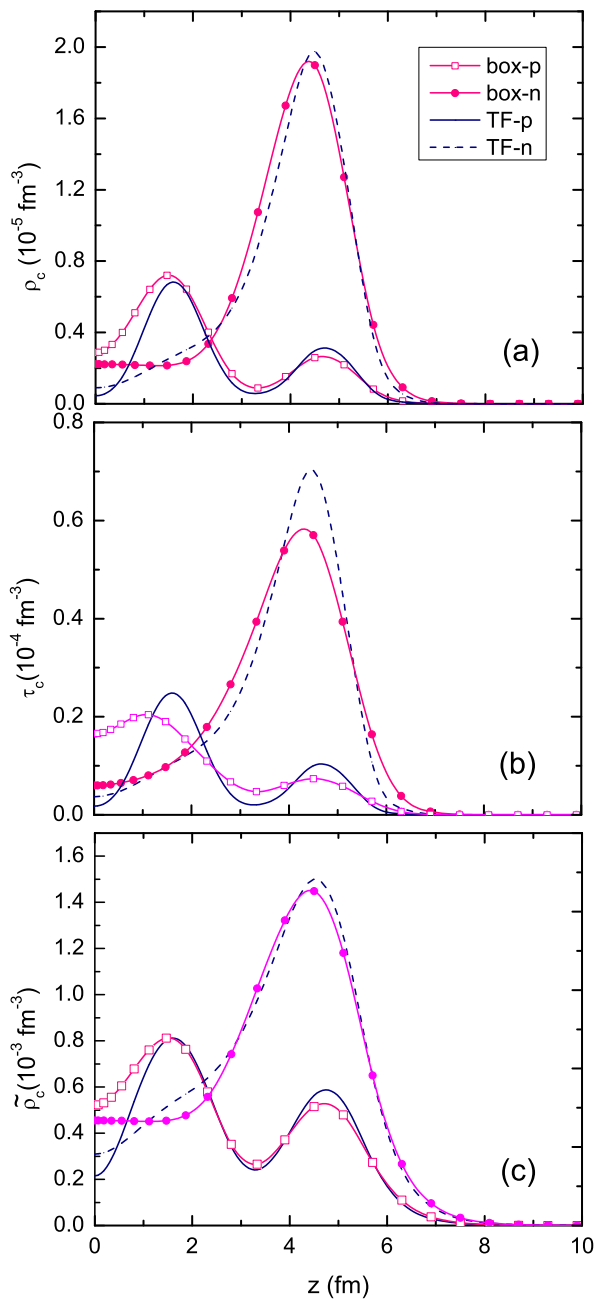}}
\caption{\label{cont-densit} (Color online) Continuum contributions
in the energy range from $E_c$=40 MeV to $E_m$=60 MeV
to (a) particle, (b) kinetic, and (c) pairing densities (see Eqs.
(\ref{densit-c}-\ref{paird-c})) of $^{70}$Zn obtained from
the TF approximation and discretized box solutions
(symbols) of {\hfbax} with the SkM* energy density functional and mixed pairing interaction.
See text for details.}
\end{figure}

To examine the quality of the TF approximation for high-energy continuum states, we compared
it with the results of box discretization calculations obtained with
the HFB solver {\hfbax} \cite{[Pei08]} for $^{70}$Zn
with the SkM$^{*}$ energy density functional (EDF) \cite{[Bar82]} and  mixed pairing
interaction. The pairing strength is taken to be $V_0=-$234.85 MeV
fm$^{-3}$ that is adjusted so as to reproduce the neutron pairing
gap of $^{120}$Sn. The calculations have been carried out with the B-spline order
of 12; the mesh size  0.6 fm, and the box size 24$\times$24
fm. The nucleus $^{70}$Zn is predicted to be spherical and has non-vanishing
pairing in both protons and neutrons.

Figure~{\ref{cont-densit}} displays the continuum contributions to particle,
kinetic, and pairing densities due to unbound states from
$E_c=$40 MeV to $E_m=$60 MeV.
(In this energy window, the discretized HFB continuum contains no deep-hole resonances.)
They were obtained from
Eqs.~(\ref{cdensit}) and from discretized {\hfbax} solutions.
It can be seen that the two methods produce very close
continuum contributions to the local densities. For the neutrons, the continuum densities are mainly
concentrated at the nuclear surface. The
proton densities  have a more pronounced volume character. We
found that this difference is mainly due to  different pairing
potentials $\Delta(\rr)$ and depends weakly on mean-field potentials
$V(\rr)$. The  continuum kinetic  and
pairing densities  have similar shapes to the continuum
particle densities. It can be seen, however, that the continuum contributions
to pairing densities are larger than  to continuum particle
densities by two orders of magnitude. This is to be expected:
the HFB continuum  is strongly affected by the pairing channel
\cite{dobaczewski96}. Similar conclusions have
been obtained in Ref.~\cite{oba}.

It is to be noted that the kinetic and pairing density integrals in Eq.~(\ref{cdensit}) are
divergent and a finite upper limit for the integration $E_c$ must be taken. In practice, the dependence on $E_c$ can be avoided
by adopting  the pairing
regularization (\ref{pair-reg}). As discussed in
Refs.~\cite{[Bul02],[Yu03],[Bor06]}, results obtained with regularized pairing
are independent on the cutoff energy $E_c$ and are very close to the results of a pairing renormalization procedure adopted in this work. Moreover, as also pointed out
in Refs.~\cite{marini,[Pap99w]}, the kinetic energy density $\tau_c$ has the same type of divergence as the pairing
density $\tilde{\rho}_c$, and the sum
of kinetic and regularized pairing energies  converges.

\subsection{The hybrid HFB strategy}

\begin{figure}[htb]
\centerline{\includegraphics[trim=0cm 0cm 0cm
0cm,width=0.43\textwidth,clip]{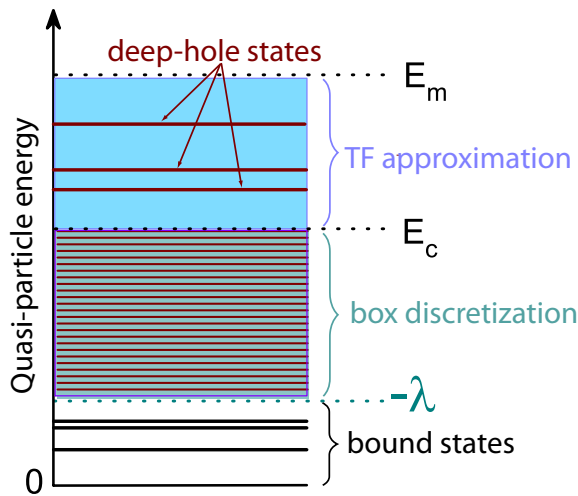}}
\caption{\label{Ehfb} (Color online) Schematic picture of the HFB quasiparticle spectrum. In the energy region $E_i\leqslant-\lambda$, quasi-particle states are bound. In the hybrid HFB strategy, the quasi-particle continuum with $E_i > -\lambda$ is divided into the low-energy continuum with  $-\lambda<E_i\leqslant E_c$, which is treated by means of the box discretization, and the high-energy continuum with $E_c<E_i\leqslant E_m$ that consists of non-resonant continuum treated by means of the TF approximation and several deep-hole states. See text for details.}
\end{figure}

We have seen that the Thomas-Fermi approximation to the non-resonant continuum
gives results very close to   the box discretization approach. This suggests a hybrid HFB strategy to separately treat
the deep-hole resonances and the high-energy non-resonant continuum. Figure~\ref{Ehfb} schematically displays the HFB quasiparticle spectrum.
As discussed earlier, the bound HFB solutions exist only in the energy region $E_i\leqslant-\lambda$.
The quasi-particle continuum with $E_i > -\lambda$ consists of non-resonant continuum and quasi-particle resonances. Among the latter ones, the deep-hole states play a distinct role. In the absence of pairing, a deep-hole excitation with energy $E_i$ corresponds to an occupied  HF state with energy
$\varepsilon_i=-E_i$. If pairing is present, it  generates a coupling of this  state with  unbound particle states with $\varepsilon_i\approx -E_i$ that gives rise to a  quasi-particle resonance with a finite width.

The low-energy continuum with  $-\lambda<E_i\leqslant E_c$ consists of many resonances that have to be computed as precisely as possible. Therefore, to this region we apply the box discretization technique. The
high-energy continuum with $E_c<E_i\leqslant E_m$ can be divided into the non-resonant part and several deep-hole states. While the non-resonant continuum can be integrated out by means of the TF approximation,  deep-hole states have to be treated separately.
Indeed, as discussed in Sec.~\ref{HFBres} below, deep-hole resonances are sufficiently narrow to be considered as a separate group of states.

To this end, we diagonalize the HF Hamiltonian,
\begin{equation}{\label{hf1}}
(h-\lambda)v^{\rm HF}_i(\rr) = \varepsilon_i v^{\rm HF}_i(\rr),
\end{equation}
to obtain wave functions $v^{\rm HF}_i(\rr)$ and single-particle energies $\varepsilon_i$ of deep-hole states. The
function $v^{\rm HF}_i(\rr)$ is a very good  approximation to the lower
HFB component  $v_i(\rr)$. In the BCS approximation, assuming the state-dependent pairing gap
($\Delta_i$ \cite{dobaczewski96,bender00}),
\begin{equation}
\Delta_i = \int \Delta(\rr)[v^{\rm HF}_i(\rr)]^2d{\rr},
\end{equation}
the effective BCS occupation factor  is:
\begin{equation}
v_k^2 = \displaystyle \frac{1}{2}\left(1-\frac{\varepsilon_k}{\sqrt{(\varepsilon_k)^2+\Delta_k^2}}\right).
\end{equation}
Consequently, $v^{\rm HF}_i(\rr)$ is normalized to the BCS occupation  $v_i^2$. The
corresponding quasi-particle energy is related to the
single-particle energy through $E_i=-\varepsilon_i$.

The hybrid HFB strategy is based on  combining the box solutions below
some low energy cutoff $E_c$, the deep-hole solutions, and high-energy TF continuum. In this way,  the total particle density (as well as other HFB densities) of the even-even system can be split into three parts:
\begin{equation}
\rho(\rr) = \sum\limits_{E_i>0}^{E_c}
2|v_i(\rr)|^2+\sum\limits_{E_{dh}>E_c}^{E_m}
\!\!\!2|v_{dh}^{\rm HF}(\rr)|^2 +\rho_{\rm c}(\rr),
\end{equation}
where the continuum particle density $\rho_{\rm c}$ of states between
$E_c$ and $E_m$ is given by Eq.~(\ref{cdensit}). In practical applications, the maximum cutoff energy $E_m$ is usually taken as
60\,MeV$-\lambda$. The  continuum contribution to the particle number,
\begin{equation}
N_c= \int \rho_c(\rr)d\rr,
\end{equation}
is always included to meet the particle number  equation.

\begin{figure}[t]
\centerline{\includegraphics[trim=0cm 0cm 0cm
0cm,width=0.44\textwidth,clip]{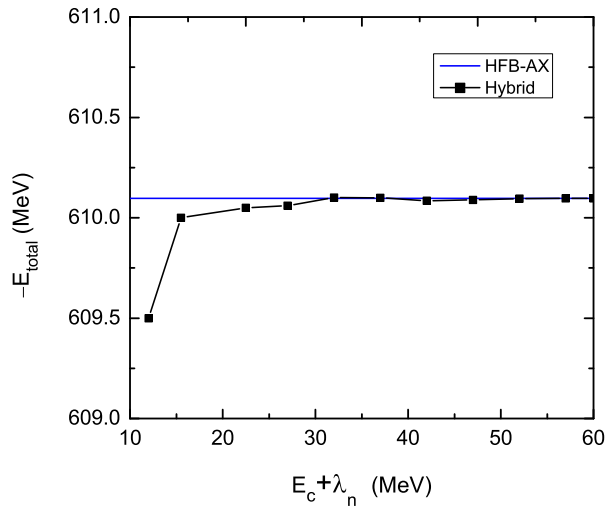}
\hspace{15pt}}
\caption{\label{cutoff} The  binding energy
of $^{70}$Zn as a function of the low-energy quasi-particle neutron cutoff $E_c$ in  the hybrid
HFB method. The HFB-AX value corresponds to $E_c$=60\,MeV+$\lambda$.}
\end{figure}

We tested this hybrid HFB strategy to calculate the binding energy
of $^{70}$Zn at different cutoff values  $E_c$. The HF equations (\ref{hf1})
are solved using {\hfbax}. Generally, the
deep-hole single-particle energies $\varepsilon_i$ obtained from
Eq.~(\ref{hf1}) are very close to the hole-like solutions of the full HFB
diagonalization. In $^{70}$Zn, for example, the quasi-particle
energy of the  1s$_{1/2}$ neutron is $E_i$=38.071 MeV, while the corresponding HF single-particle energy is $-\varepsilon_i$=38.055 MeV.

Figure~\ref{cutoff} shows the total binding energy
 as a function of the neutron cutoff  $E_c$.
It is seen that the total energy is perfectly stable for $E_c+\lambda_n$$>$30\,MeV,
and it is equal to the binding energy obtained by means of the full bock discretization ($E_c=E_m$). At a low value of $E_c+\lambda_n$$=$15\,MeV,
the error of the hybrid method is only about 100\,keV.
At even lower values of cutoff, the TF approximation deteriorates
rapidly \cite{[Bor06]}. Note that the pairing strength in the hybrid HFB should not be renormalized with $E_c$. The choice of the cutoff $E_c$ is determined by positions of  deep-hole levels; this information can be obtained  by solving the HF problem.
The hybrid strategy can also be modified to work with
the pairing regularization procedure~\cite{[Bor06]}. However, as seen in Fig.~\ref{cutoff},  this technique
cannot be used with a very low cutoff $E_c$ where resonances are
densely populated.

\section{Properties of deep-hole resonances of HFB}\label{HFBres}

In this section, we study several approximate methods to calculate
widths of HFB resonances. Calculations are performed for  $^{70}$Zn
using the SkM$^{*}$ EDF and the mixed pairing interaction.

\subsection{The smoothing and fitting method}

\begin{figure}[htb]

\centerline{\includegraphics[trim=0cm 0cm 0cm
0cm,width=0.45\textwidth,clip]{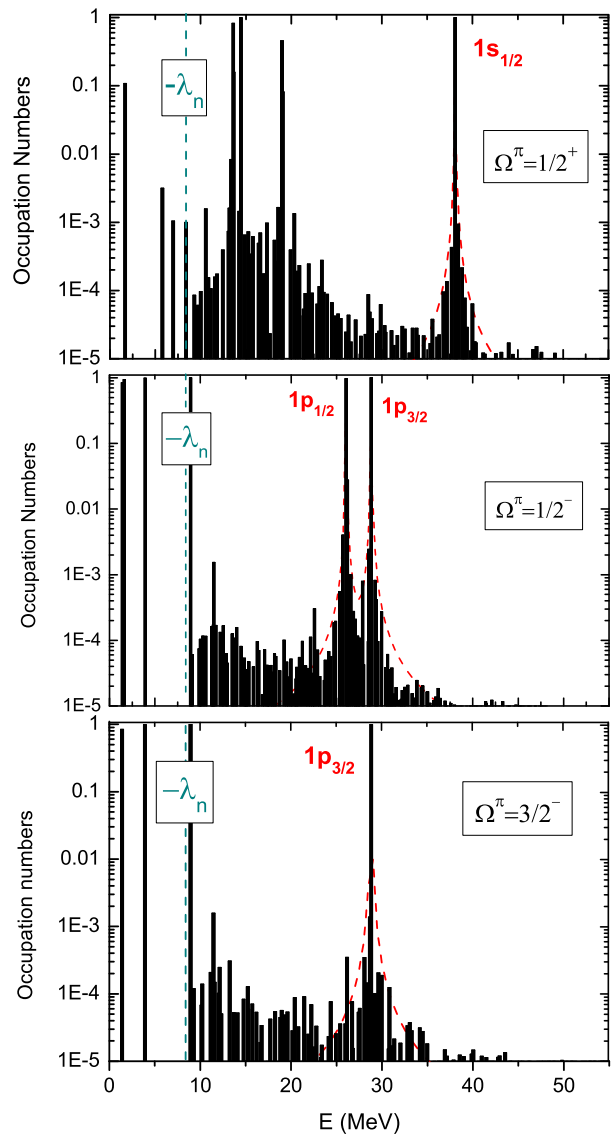}}
\caption{\label{smoothd} (Color online) Occupation numbers of
the discretized neutron quasi-particle continuum states with $\Omega^{\pi}=1/2^{+}$ (top),   $\Omega^{\pi}=1/2^{-}$ (middle),  and  $\Omega^{\pi}=3/2^{-}$ (bottom) calculated for $^{70}$Zn  with {\hfbax}. Since $^{70}$Zn is spherical,
 angular momenta of discretized resonance states are marked.
The corresponding  Breit-Wigner envelopes are  indicated by dashed lines.
Note that the envelopes for $1p_{3/2}$ magnetic substates are identical.}
\end{figure}

From the box discretization of {\hfbax}, one obtains a finite and very
dense spectrum of continuum states. The deep-hole states are no longer isolated as in the BCS approximation; they become fragmented due to the pairing coupling with the neighboring particle-like continuum and  acquire a decay width. The energy distribution of the
occupation numbers
\begin{equation}\label{vocc}
v_i^2= \int |v_i(\rr)|^2\,d\rr
\end{equation}
has roughly the Breit-Wigner shape. Figure~\ref{smoothd}
displays occupation probabilities for the $\Omega^{\pi}$=$1/2^{+}$, $1/2^{-}$,
and $3/2^{-}$ discretized neutron quasi-particle states in $^{70}$Zn
as a function of quasi-particle energy $E_i$.
Note that the angular momentum projection $\Omega$ and
parity $\pi$ are good quantum numbers in {\hfbax}.  It is apparent that the $v_i^2$ distributions
of $1s_{1/2}$, $1p_{3/2}$, and $1p_{1/2}$ deep-hole states have resonance-like structure. For spherical
nuclei, degenerate resonances with different values of $|\Omega|$ belonging to the same shell  are expected to
have the same width, and this is indeed the case for both magnetic substates of $1p_{3/2}$.
Breit-Wigner resonance-like structures in occupation probabilities have
also been predicted by the recent continuum HFB calculations~\cite{oba}. In
the Green function approach~\cite{Belyaev,oba}, the occupation
probability is related to the continuum level density, which
corresponds to a Breit-Wigner shape for resonances~\cite{kruppa99}.
In the present work, we propose a straightforward smoothing and
fitting method to estimate  resonance widths from
discrete $v_i^2$ distributions.

To extract resonance parameters from the discrete distribution of
$v_i^2$, we first smooth it using a Lorentzian shape function
$w(x)=1/[2\pi(x^2+1/4)]$~\cite{arai,kruppa99}. The occupation
numbers $v_i^2$ of states above the Fermi level are smoothed out by means of folding with $w(E/\Gamma)$,
\begin{equation}\label{vsmoothed}
\overline{v^2}(E)=
\sum_i\frac{v_i^2}{\Gamma}w\Big(\frac{E-E_i}{\Gamma}\Big),
\end{equation}
where $\Gamma$ is a smoothing parameter.

The smoothed Breit-Wigner distribution is given by a compact expression~\cite{arai,kruppa99}:
\begin{equation}\label{vfit}
\begin{array}{ll}
\overline{v^2_R}(E) =& \displaystyle
\frac{\Gamma_r\beta}{2\pi^2}\Big\{(E-E_r)^2-\frac{(\Gamma^2-\Gamma_r^2)}{4}\Big\}\vspace{5pt}\\
\vspace{5pt}
&\displaystyle\times\Big\{\frac{\pi}{2}+\tan^{-1}\big(\frac{2E}{\Gamma}\big)\Big\}
\displaystyle +\frac{\Gamma\beta}{2\pi^2}\\ \vspace{5pt} &
\displaystyle\times\Big\{(E-E_r)^2+\frac{(\Gamma^2-\Gamma_r^2)}{4}\Big\}\\\vspace{5pt}
&
\displaystyle\times\Big\{\frac{\pi}{2}+\tan^{-1}\big(\frac{2E_r}{\Gamma}\big)\Big\}\\
&\displaystyle+\frac{\Gamma_r\Gamma}{4\pi^2}(E-E_r)\beta\ln\Big[\frac{4E^2+\Gamma^2}{4E_r^2+\Gamma_r^2}\Big],
\end{array}
\end{equation}
where\begin{equation}
\beta=\big[(E-E_r)^2+(\Gamma-\Gamma_r)^2/4\big]^{-1}\big[(E-E_r)^2+(\Gamma+\Gamma_r)^2/4\big]^{-1}.
\end{equation}

In the smoothing and fitting method, we fit the smoothed $\overline{v^2}(E)$ distribution (\ref{vsmoothed}) using a two-term
expression:
\begin{equation}{\label{smo-fit}}
\overline{v^2}(E) = v^2\overline{v^2_R}(E)+ v^2_b(E).
\end{equation}
The first term is the smoothed Breit-Wigner resonance shape. The background
contribution $v^2_b(E)$ is assumed to be a slowly changing
Fermi-Dirac function $a/(e^{-E/b}+1)$ characterized by parameters $a$ and $b$.
In order to deduce the resonance energy $E_r$, width $\Gamma_r$, and the
occupation number $v^2$($v^2\leqslant1$), the method of least
squares is used.

\begin{figure}[t]
\centerline{\includegraphics[trim=0cm 0cm 0cm
0cm,width=0.46\textwidth,clip]{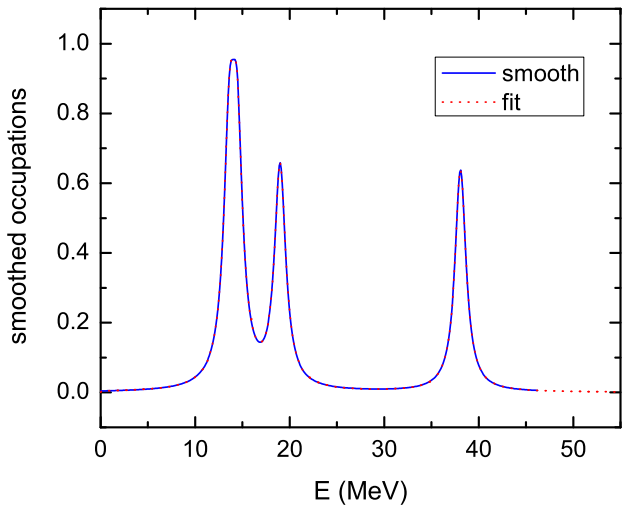}}
\caption{\label{smooth} (Color online) Smoothed occupation
numbers (\ref{vsmoothed}) and the corresponding smoothed Breit-Wigner distribution (\ref{vfit}) for the $\Omega^{\pi}=1/2^{+}$
neutron HFB resonances in $^{70}$Zn.}
\end{figure}
%%%%%%%%%%%%%%%%%%%%%%%%
The results of such a procedure for the $\Omega^{\pi}=1/2^{+}$
neutron HFB resonances in $^{70}$Zn are illustrated in Fig.~\ref{smooth}: the fitted
Breit-Wigner resonances agree very well with the smoothed occupation
numbers. This demonstrates that the  discretized
occupation numbers  have a Breit-Wigner shape. One has to note, however,
that for threshold resonances (such as those close to $E=-\lambda$), the
distribution can strongly deviate from Breit-Wigner  \cite{Grasso}.

As it has been pointed out in Ref.~\cite{kruppa99}, the fitting
curve could be dependent on the smoothing parameter $\Gamma$. We
checked that such a dependence is not significant as far as the
resonance width is concerned if the value of  $\Gamma$ around 0.8
MeV is taken. Generally, one should use $\Gamma\gg\Gamma_r$.  A
very small smoothing parameter is not sufficient to smooth
out the finite discretization effects. On the other hand, a large smoothing
parameter can underestimate the resonance width~\cite{kruppa99}.
Numerical errors can also arise from the fitting procedure if several
resonances are overlapping. The discretization can yield a  crude representation
of occupation distribution   if too small a box
is used ~\cite{oba}. A high discretization
resolution is of particular importance when it comes to
narrow resonances. Another advantage of the direct smoothing-fitting
method discussed here is that it can  be applied to resonances in deformed
nuclei.

\subsection{The box stabilization method}

The stabilization method \cite{hazi} is an $\mathcal{L}^2$ method, and has been
used to obtain precisely the resonance energy $E_r$ and widths
$\Gamma_r$ in atomic \cite{mandelshtam94,ryaboy,Salzgeber}
and nuclear \cite{zhang1,zhou} physics. Based on
the box solutions, the HFB resonances are expected to be localized
solutions with energies weakly affected by changes of the box size.
The stabilization method allows to obtain the resonance parameters from the
box-size dependence of quasi-particle eigenvalues.

To this end, one introduces the continuum
level density $\Delta(E)$,
\begin{equation}
g(E)={\rm Tr}[\delta(E-H)-\delta(E-H_0)],
\end{equation}
where $H_0$ is the non-interacting Hamiltonian of the system. The continuum
level density is related to the phase shift by
\begin{equation}
\delta(E)=\pi\int_0^E g(E')\,dE'.
\end{equation}

In a modified stabilization method, one can obtain the phase shift
using the box-discretized continuum~\cite{mandelshtam94}.
Firstly, we compute  quasi-particle energies $E_j(L)$ with different
box size ${L}$ using the spherical HFB solver {\hfbrad} \cite{HFBRAD}.
%%%%%%%%%%
\begin{figure}[htb]
\centerline{\includegraphics[trim=0cm 0cm 0cm
0cm,width=0.48\textwidth,clip]{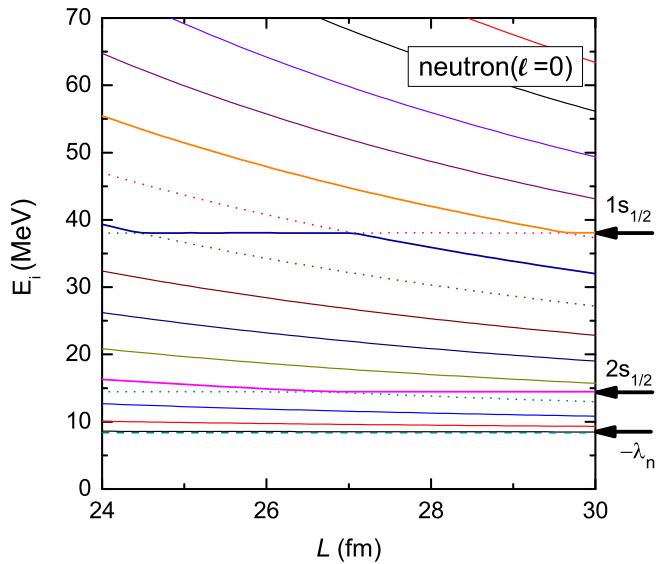}}
\caption{\label{levels} (Color online) The neutron discretized quasi-particle $\ell$=0
spectrum  of $^{70}$Zn as a function of the box
size $L$. Calculations have been carried out with the spherical
HFB solver {\hfbrad}~\cite{HFBRAD}. Unlike the non-resonant box continuum,  resonances   are practically $L$-independent.}
\end{figure}
%%%%%%%%%%%%
Figure~{\ref{levels}} shows that the discretized quasi-particle energies generally smoothly
decrease as the box size increases. Around the resonances, however,
the energies are fairly constant. Starting from a
sufficiently large box  $L_0$, calculations are done on a grid with the spacing
 $\delta L$:
\begin{equation}{\label{steps}}
L = L_0+\Delta L, ~~~\Delta L = (M-1)\delta L.
\end{equation}
In this way, eigenvalues $E_j$ are stored in arrays $E_j(L)$. By
using the Akima interpolation~\cite{akima}, we obtain the values of  $L_j(E)$
 corresponding to a box size with which the $j^{\rm th}$
eigenvalue equals to $E$.

In the next step,  the phase shift is obtained from~\cite{mandelshtam94}:
\begin{equation}{\label{phases}}
\delta(E)= \pi N(E)+\frac{\pi}{\Delta L}\sum_j(L_0+\Delta L-L_j(E)),
\end{equation}
where $N(E)$ is the number of eigenvalues for which $E_j(L_0)<E$.
To obtain the resonance energy $E_r$ and the
width $\Gamma_r$,
we fit the phase shift $\delta(E)$ by using the expression:
\begin{equation}
\tilde\delta(E)=\arctan\left(\frac{2(E-E_r)}{\Gamma_r}\right)+\tilde\delta_b(E),
\end{equation}
where $\tilde\delta_b(E)$ represents the smoothly changing background
contribution parametrized as:
\begin{equation}{\label{bk}}
\tilde\delta_b(E) = a + b(E-E_r).
\end{equation}

\begin{figure}[htb]
\centerline{\includegraphics[trim=0cm 0cm 0cm
0cm,width=0.46\textwidth,clip]{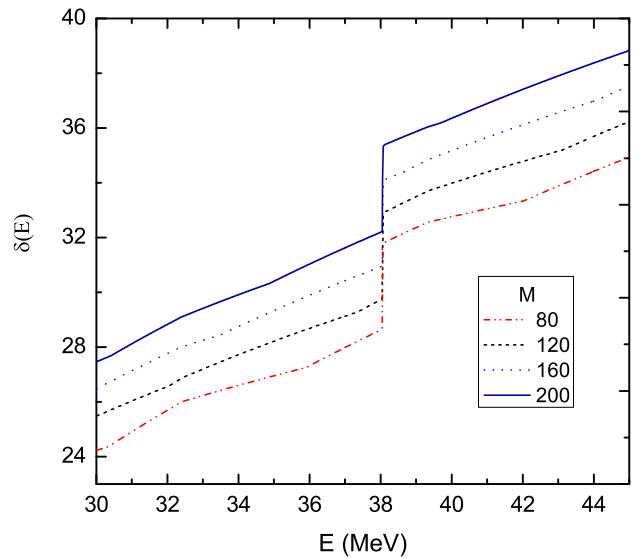}}
\caption{\label{msteps} The  phase shifts $\delta(E)$ of the
neutron 1s$_{1/2}$ state of $^{70}$Zn obtained with the stabilization method
with different iteration steps (\ref{steps}). Here,  $\delta L$=0.06 fm and
$L_0$=24 fm.}
\end{figure}
%%%%%%%
Figure~{\ref{msteps}} shows the phase shifts of the neutron 1s$_{1/2}$
resonance obtained with different step numbers $M$. It can be seen that the
background becomes more smooth as the step numbers increase.
Generally, a large $\Delta L$ is necessary to smooth out the background,
so as to reduce the fitting errors in $a$ and $b$  in
Eq.~({\ref{bk}}). In our calculation, we also find that for very
narrow resonances, a small value of $\delta L$ is  important for the
interpolation precision. In principle, as
\begin{equation}
\delta L\rightarrow 0 ~~~{\rm and}~~~ \Delta L \rightarrow \infty,
\end{equation}
one can get very accurate resonance parameters. In practice, however,
calculations with a large box and small step sizes are very
expensive even with a fast {\hfbrad} solver. The eigenvalue spectrum of {\hfbrad} is  sparse, resulting in statistical errors in the
calculations of phase shift. For low-energy HFB
resonances, this situation is even worse since these eigenvalues
change very slowly as the box size increases, as shown in
Fig.~{\ref{levels}}. In this case, the calculated background is not
sufficiently smooth.
Currently, the stabilization method is limited to spherical
boxes as in {\hfbrad}. The extension of the stabilization method to deformed cases will be the subject of future work.

\subsection{The perturbative expression}

Assuming that the pairing coupling
can be treated perturbatively, the width of
a deep-hole HFB resonance
can be given by the Fermi Golden rule
~\cite{bulgac80,Belyaev,dobaczewski96}:
\begin{equation}{\label{overlap}}
\Gamma=2\pi|\langle u_{0E}|\Delta|v^{\rm HF}| \rangle|^2,
\end{equation}
where $\Delta$ is the pairing potential; $v^{\rm HF}$ is the bound  wave function of the HF potential corresponding to the single-particle energy $-E+\lambda$, and
$u_{0E}$ is  the scattering solution of the
HF equation with energy $E+\lambda$.

At a large distance, the  scattering wave function $u_{0E}$ has the  usual asymptotic form:
\begin{equation}
u_{0E} ={1\over r} \sqrt{\frac{2m}{\hbar^2\pi k}}\left(\cos\delta_lF_l(\eta,
kr)+\sin\delta_lG_l(\eta, kr)\right),
\end{equation}
where $F_l$ and $G_l$ are the regular and irregular Coulomb
functions, respectively; $\delta_l$ is the phase shift; and
$k=\sqrt{2m(E+\lambda)/\hbar^2}$.  The scattering wave function is
found with the help of Ref.~\cite{papp}.  In the calculation of
scattering wave functions, $\hbar^2/2m$ is taken as a
constant 20.734 MeV, to avoid problems related to  the density-dependent effective mass.

\begin{table*}[t]
 \caption{\label{tab-widths}
Energies (in MeV) and widths (in keV) of deep-hole HFB resonances in $^{70}$Zn calculated with the box stabilization method (box),
 smoothing-fitting method (smf), and
perturbative expression (per). See text
 for more details.}
\begin{tabular}{|c|cccc|cccc|} \hline
  & \multicolumn{4}{c|}{Neutron }& \multicolumn{4}{c|}{Proton } \\
~~ states~~ & ~~~~~E~~~~~ &  ~~~$\Gamma_{\rm box}$~~~&~~~
$\Gamma_{\rm  smf}$ ~~~&~~~$\Gamma_{\rm per}$
  ~~~&~~~~~E ~~~~~&~~~$\Gamma_{\rm box}$~~~&~~~ $\Gamma_{\rm smf}$ ~~~&~~~
$\Gamma_{\rm per}$~~
\\ \hline \vspace{1pt}
$1s_{1/2}$&38.075&1.23&0.80 &0.98 &35.147&0.64& 0.54 &0.40 \vspace{1pt} \\
$1p_{3/2}$& 28.801&3.42&4.1&0.04&25.424&3.22&2.04&2.27 \vspace{1pt}
\\
$1p_{1/2}$& 26.043&6.01 &7.0&0.19&22.851&2.04&1.57&1.26  \vspace{1pt}\\

$1d_{5/2}$&18.987&13.1&10.6 & 22.7 &15.320&-&0.12 &0.09 \\
\hline
\end{tabular}
\end{table*}

Table {\ref{tab-widths}} displays the widths of high-energy
deep-hole states in $^{70}$Zn  calculated with the smoothing-fitting method,
 box stabilization method, and  perturbative expression.
 It is seen that the widths obtained
with the smoothing-fitting method and  box stabilization method
are always fairly close. For protons, all three methods yield similar results. However, the perturbative expression
predicts widths that are too small for the 1p$_{3/2}$ and 1p$_{1/2}$ neutron
states.  The difference is due to
a surface-peaked neutron pairing potential that is very
different from that of protons (see Fig.\ref{cont-densit}). Indeed,
for surface-peaked pairing,
the overlap between $u_{0E}$ (which exhibits rapid spatial oscillations), $v^{\rm HF}$, and $\Delta(r)$ is too sensitive to small changes
in $u_{0E}$. For more discussion on limitations of the
perturbative expression (\ref{overlap}), see also the schematic model  analysis in Ref.~\cite{Belyaev}.

\begin{figure}[b]
\centerline{\includegraphics[trim=0cm 0cm 0cm
0cm,width=0.46\textwidth,clip]{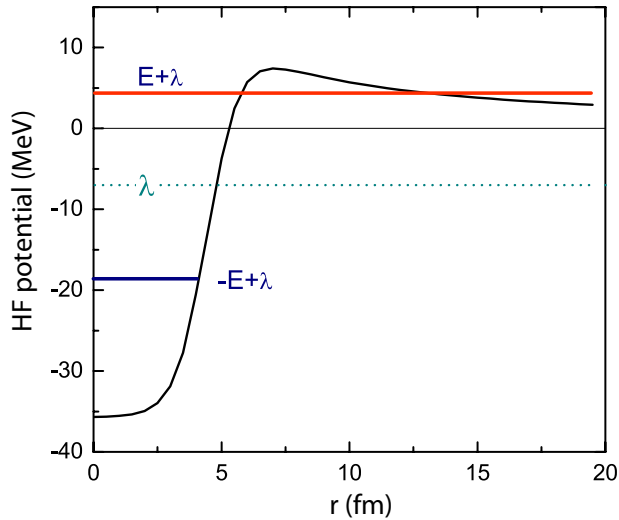}}
\caption{\label{special-hfb} (Color online) A special kind of
deep-hole HFB proton resonance, in which the upper component of the HFB
wave function is quasi bound, due to the confining effect of the
Coulomb-plus-centrifugal barrier.}
\end{figure}
%%%%%%%%%%%%%%
Compared to
other deep-hole resonances in Table~{\ref{tab-widths}}, the width of the proton 1d$_{5/2}$ state is very small. This is because the upper component wave function of this $\ell=2$ state is quasi bound, due to the confining effect of the  Coulomb-plus-centrifugal barrier (see
Fig.~{\ref{special-hfb}}). For such narrow resonances, the stabilization
method cannot easily be applied.

\section{applications to weakly bound nuclei}\label{examples}

The effects of continuum  are expected to increase when moving towards the neutron  drip line~\cite{[Dob97a]}. Especially for very weakly bound nuclei
exhibiting halo structures, there exists a strong interplay between pairing and
continuum~\cite{[Ben00],[Yam05],[Hag11]}. Here, we
investigate the role of continuum contributions in the neutron-rich Ni isotopes. These nuclei have been studied by the Gamow
HFB (GHFB) method and the exact quasi-particle resonance widths of
$^{90}$Ni are available~\cite{Michel}. The GHFB work provides an excellent benchmark for our approximate resonance widths.

\begin{figure}[b]
\centerline{\includegraphics[trim=0cm 0cm 0cm
0cm,width=0.46\textwidth,clip]{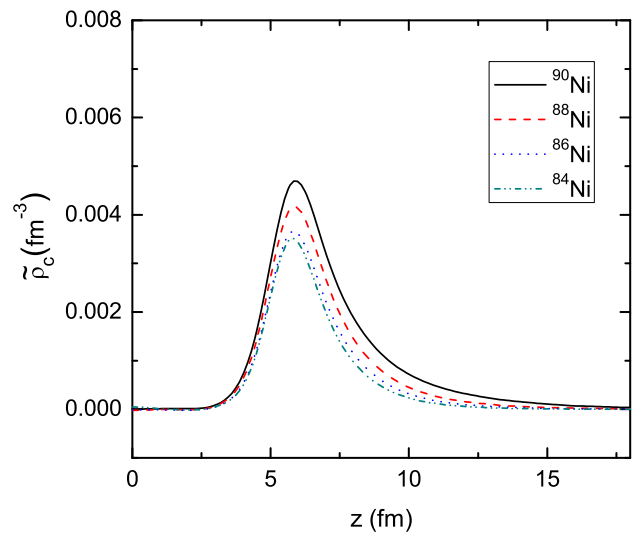}}
\caption{\label{ni-iso} (Color online) The continuum contributions
to the neutron pairing densities of neutron-rich $^{84,86,88,90}$Ni isotopes, calculated  with
Eq.~(\ref{cdensit}) with $E_c$=30 MeV and $E_m$=60 MeV and mixed pairing interaction.
See text for details.}
\end{figure}
%%%%%%
As in Ref.~\cite{Michel}, the calculations have been carried out with the
SLy4 EDF \cite{sly4} and the surface pairing interaction \cite{[Dob02c]}, with the
strength  $V_0$=$-$519.9 MeV fm$^{-3}$.  In Fig.\ref{ni-iso},
we show  the high-energy continuum contributions to the neutron
pairing densities in $^{84,86,88,90}$Ni, obtained using the
TF approximation (\ref{cdensit}). It can be seen that the continuum
pairing density acts  in the surface region, and increases as
nuclei become less bound. This is consistent with the earlier findings \cite{dobaczewski96} that  continuum  becomes important for nuclei close to the drip line.

\begin{figure}[htb]
\centerline{\includegraphics[trim=0cm 0cm 0cm
0cm,width=0.46\textwidth,clip]{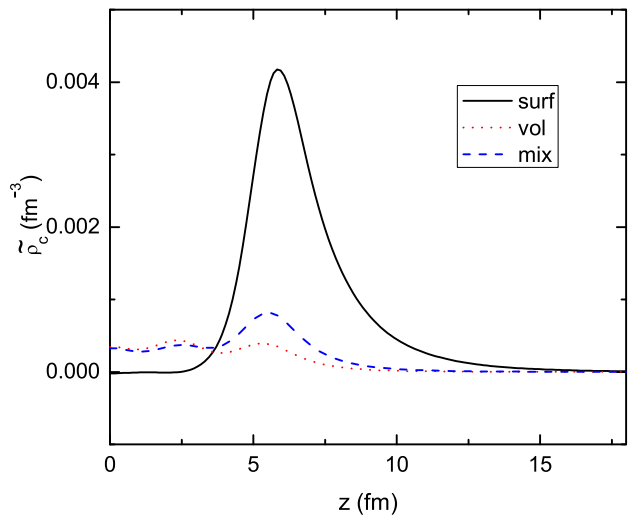}}
\caption{\label{ni88} (Color online) Similar to
Fig.~{\ref{ni-iso}} but with the volume, surface, and mixed pairing
interaction in $^{88}$Ni. }
\end{figure}

Figure~\ref{ni88} illustrates the interaction dependence of the pairing-continuum effect. Namely, it displays the neutron continuum
pairing densities for $^{88}$Ni with the
volume pairing ($V_0$=$-$185.026 MeV fm$^{-3}$), surface pairing as in Fig.\ref{ni-iso}, and
mixed pairing ($V_0$=$-$284.36 MeV fm$^{-3}$). (The
pairing strengths $V_0$ have all been adjusted to reproduce the neutron
pairing gap of $^{120}$Sn.) It is seen that the high-energy continuum contribution to
pairing density strongly depends on character of
pairing interaction \cite{Belyaev,dobaczewski96}.
In particular, in the case of
surface pairing, the continuum contribution is remarkably larger
than for other two pairing functionals, indicating its very different behavior
in weakly bound nuclei \cite{[Dob01a]}. Actually, as it has been discussed~\cite{[Pei08]} the surface
pairing is essential for the existence of bound  $^{90}$Ni.

\begin{figure}[htb]
\centerline{\includegraphics[trim=0cm 0cm 0cm
0cm,width=0.46\textwidth,clip]{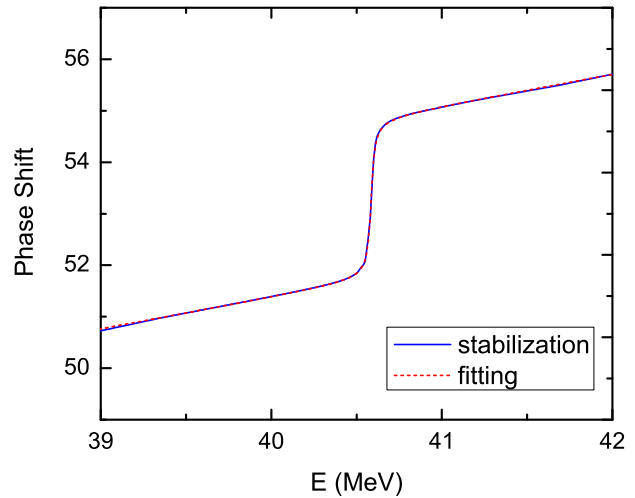}}
\caption{\label{ni90-fit} (Color online) The phase
shift of the neutron 1p$_{3/2}$ resonance in $^{90}$Ni using  Eqs.~(\ref{bk})
and (\ref{phases}). }
\end{figure}

\begin{table*}[htb]
 \caption{\label{tab-ni90}
Energies (in MeV) and widths (in keV) of HFB neutron resonances
 in $^{90}$Ni,
 calculated using the box stabilization  (box)
 and  smoothing-fitting  (smf) methods. They are
  compared to the GHFB solutions of Ref.~\cite{Michel}.}
\begin{tabular}{|c|cccc|} \hline
~~~~~~ states~~~~~~ & ~~~~~~~~~$E_r$~~~~~ ~~~~&~~~~ ~~~$\Gamma_{\rm
box}$~~~~~~~&~~~~~ $\Gamma_{\rm smf}$ ~~~~~&~~
~~~$\Gamma_{\rm GHFB}$~~~~~~~
\\ \hline \vspace{1pt}
$1s_{1/2}$&51.419&-&1.1e-3 &1.09e-3 \vspace{1pt} \\
$1p_{3/2}$&40.588&30.84 & 20.17&  27.28 \vspace{1pt}\\
$1p_{1/2}$&38.770&32.26 & 34.67 &  27.14 \vspace{1pt}\\
$1d_{5/2}$&29.039&1.31&1.37&  0.78 \vspace{1pt}\\
$1d_{3/2}$&25.017&25.44&23.08&  22.57 \vspace{1pt}\\
$2s_{1/2}$&24.319&50.36&40.87&  46.00 \\
$1f_{7/2}$&17.554&401.79&413.04 & 397.37 \\
$2p_{3/2}$&12.538&499.98&471.22 & 490.56 \\
$1f_{5/2}$&10.981&672.44&651.12 & 645.64 \\
$2p_{1/2}$&10.816&440.76&376.19 & 404.30 \\
$1g_{9/2}$&6.519&1.56&0.52&  0.81 \\
$2d_{5/2}$&3.270&221.08&105.17 & 194.18 \\
$2d_{3/2}$&2.310&611.50&643.88 & 560.61 \\
$1g_{7/2}$&3.348&69.09&75.13&  63.61 \\
$1h_{11/2}$&5.527&162.46&173.66 & 131.78 \\
\hline
\end{tabular}
\end{table*}
%%%%%%%%%%%%%%%%%%%%%%%%%%%
Table \ref{tab-ni90} displays the widths of neutron
resonances in $^{90}$Ni calculated using  different techniques. In the
calculations with the box stabilization method, we used
$L_0$=22 fm, $M$=400, and $\delta L$=0.06 fm, see Eq.~ (\ref{steps}). Figure~\ref{ni90-fit} illustrates the quality of calculations for the neutron 1p$_{3/2}$ resonance.
It is seen that the fitted phase shift agrees well with that
obtained from the stabilization method. Systematically, the box stabilization method
predicts slightly larger widths as compared to  GHFB.
This is consistent with findings of
Ref.~{\cite{ryaboy}} where the stabilization method  generally
overestimates the widths by 10$\%$. In particular, it is shown that
the widths of very narrow resonances are largely overestimated. The
width of the 1s$_{1/2}$ state is so narrow that it is beyond the
applicability of the stabilization method. The very narrow 1g$_{9/2}$ state  belongs to the class of special
HFB resonances of  Fig.\ref{special-hfb}. Among the
resonances, the 1g$_{7/2}$ and 1h$_{11/2}$ states are also
single-particle resonances in the Hartree-Fock theory~\cite{Michel}.
Other than that,  Table \ref{tab-ni90} demonstrates  that the stabilization method works  well for all the HFB resonances except for extremely narrow
ones.
%%%%%%%%%%%%%%%%%%%%%%%%%%

Within the smoothing-fitting method, the quasi-particle energy
spectrum is obtained by {\hfbax} by taking a  large box of
38$\times$38 fm. The widths given by the smoothing-fitting method
agree with the exact numbers within a factor of two. We
have found that the fitting precision is compromised  when
several resonances overlap. For the low-energy resonances,
the total occupation numbers $v^2$ in Eq.(\ref{smo-fit}) are very
small and can induce additional fitting errors. Besides, as we have
discussed earlier, it is not proper to fit the occupation
probabilities of low-energy resonances near the Fermi energy using a
Breit-Wigner shape. In spite of all those reservations, the precision of the
smoothing-fitting method is quite satisfactory.

For neutron HFB resonance widths in Ni isotopes, our
calculations predict that generally the widths would slowly increase
as the drip line  is approached. For example, the widths of the neutron
1p$_{3/2}$ state in $^{86}$Ni, $^{88}$Ni, and $^{90}$Ni calculated by
the stabilization method are, respectively,  25.97 keV, 28.25 keV, and 30.84 keV.
This is consistent with Fig.~{\ref{ni-iso}}: the widths grow with  the
increased  pairing-continuum coupling.

\section{Conclusions}\label{conclusions}

In this work, we performed a comprehensive study of quasi-particle continuum  within the HFB theory. The purpose of this investigation is twofold. Firstly, we tested a truncation scheme based on the Thomas-Fermi approximation to limit the continuum space in realistic calculations carried out in huge configuration spaces (or large spatial boxes) that yield huge amounts of discretized unbound states.
Secondly, we studied properties of HFB resonances, including deep-hole states. We compare several methods to estimate resonance width and discuss their strengths and weaknesses.

The TF approximation to the high-energy continuum states in the hybrid HFB variant, in which deep-hole states and non-resonant  continuum are
separately treated,
has been found very effective for it fully reproduces results obtained with the full box discretization.
The high-energy non-resonant continuum  has been  found to have a similar
spatial behavior  for  particle, kinetic, and
pairing densities. This distribution is mainly determined by
the pairing potential. The continuum contribution to the pairing
density is substantial for weakly bound nuclei and it has appreciable spatial extension. The hybrid method will be useful in reducing the
computational cost of  3D coordinate-space HFB calculations.

The HFB quasi-particle resonance is unique to the HFB theory and is
a fascinating topic in its own right. We examined three approximate methods to study
the resonance widths based on HFB box solutions. In the smoothing-fitting method,
resonance parameters are obtained by fitting the smoothed occupation numbers obtained from discretized solutions. The box stabilization method is  based on the fact that quasi-particle energies
of continuum states change with the box size. By comparing with
the exact Gamow HFB results obtained by imposing outgoing boundary conditions
\cite{Michel}, we have demonstrated that
the stabilization method works fairly well for all HFB
resonances,  except for the very narrow ones.
The smoothing-fitting method is also very effective and can easily be extended to
deformed cases.
The perturbative
 Fermi golden rule  is found to be
unreliable for calculating widths of neutron resonances. The only exceptions are narrow metastable deep-hole states such as high-$\ell$ states and low-lying proton resonances.

Illustrative examples have been provided for the drip-line Ni isotopes. We found that continuum densities strongly depend on the density dependence of pairing interaction. In particular,  surface pairing produces very large continuum pairing densities.
The obtained neutron widths of
$^{90}$Ni are generally larger than that of stable nuclei. The presence of broad
quasi-particle resonances  in weakly bound nuclei suggests that
quasi-particle continuum   plays an important
role in the description of excited states. In
addition, we expect that the determination of neutron resonance
widths can be useful to estimate  neutron emission half-lives of
excited states above the neutron emission threshold.

In summary, we have demonstrated that one can implement powerful
approximations to incorporate the vast quasi-particle continuum space
without explicitly imposing scattering or outgoing boundary conditions.
We expect our work can also be useful in the context of continuum-QRPA
applications. The approximate techniques used in this study
have  demonstrated the precision of the box discretization in representing
the continuum and deep-hole resonances, especially
 important for nuclei near drip
lines where  continuum effects  are  large.

\begin{acknowledgments}
Useful discussions with J. Dobaczewski, G. Fann, R. Harrison, R. Id Betan, N. Michel, and M. Stoitsov are gratefully acknowledged. This work was supported in part
by the U.S. Department of Energy under Contract Nos.
DE-FG02-96ER40963 (University of Tennessee) and
DE-FC02-09ER41583 (UNEDF SciDAC Collaboration), and by  the Hungarian OTKA Fund No. K72357.
Computational resources were provided through an INCITE award ``Computational
Nuclear Structure" by the National Center for Computational Sciences (NCCS) and
National Institute for Computational Sciences (NICS) at Oak Ridge National Laboratory, and the National Energy Research Scientific Computing Center.
\end{acknowledgments}

%\bibliographystyle{unsrt}
%\bibliography{continuumhfb}

\end{document}